**Title:**

**The Spatial Selective Auditory Attention of Cochlear Implant Users in Different Conversational Sound Levels**

**Authors:**

**Sara Akbarzadeh[1]**

[1]Erik Jonsson School of Engineering and Computer Science, University of Texas at Dallas

800 West Campbell Road, Richardson, TX, 75080, USA.

EMAIL: Sara.Akbarzadeh@utdallas.edu

**Sungmin Lee[2]**

[2]Department od speech language pathology and audiology, Tongmyong University

428 Sinseon-ro, Nam-gu, Busan, 48520, Republic of Korea.

EMAIL: Slee18@tu.ac.kr

**Chin Tuan-Tan[3]**

[3]Erik Jonsson School of Engineering and Computer Science, University of Texas at Dallas

800 West Campbell Road, Richardson, TX, 75080, USA.

EMAIL: Chin-Tuan.Tan@utdallas.edu

**Acknowledgments of support:**

**This research was supported by STARs (Science and Technology Acquisition and Retention) program. The authors extend thanks to the subjects who participated in this study.**

# Abstract:

In multi speakers environments, cochlear implant (CI) users may attend to a target sound source in a different manner from the normal hearing (NH) individuals during a conversation. This study attempted to investigate the effect of conversational sound levels on the mechanisms adopted by CI and NH listeners in selective auditory attention and how it affects their daily conversation. Nine CI users (five bilateral, three unilateral, and one bimodal) and eight NH listeners participated in this study. The behavioral speech recognition scores were collected using a matrix sentences test and neural tracking to speech envelope was recorded using electroencephalography (EEG). Speech stimuli were presented at three different levels (75, 65, and 55 dB SPL) in the presence of two maskers from three spatially separated speakers. Different combinations of assisted/impaired hearing modes were evaluated for CI users and the outcomes were analyzed in three categories: electric hearing only, acoustic hearing only, and electric+acoustic hearing. Our results showed that increasing the conversational sound level degraded the selective auditory attention in electrical hearing. On the other hand, increasing the sound level improved the selective auditory attention for the acoustic hearing group. In NH listeners, however, increasing the sound level did not cause a significant change in the auditory attention. Our result implies that the effect of the sound level on the selective auditory attention varies depending on hearing modes and the loudness control is necessary for the ease of attending to the conversation by CI users.

# 1- Introduction

Selective auditory attention is the ability of the auditory system to attend a target sound source and ignore the competing sounds in multi-speaker environments, known as the cocktail party scenario. Spatial, temporal and frequency cues help to identify and separate the speech streams. If these cues are not accessible in the auditory pathway, this may result in decreased spatial selective auditory attention and speech intelligibility [1], [2].

Previous studies have shown that spatial selective auditory attention is degraded by hearing loss [1]–[6]. Most hearing impaired (HI) listeners have suffered from poor spatial hearing, especially when they encounter social settings where multiple people talk in a group. This is particularly the case for cochlear implant (CI) listeners, who have severe hearing loss and their performance fall much below that of the normal hearing (NH) listeners in spatial hearing tasks [7]–[10]. Beside intelligibility of the target stimulus, which is commonly used to explain the degraded auditory selective attention, localizing the target speech source among distractors can also concurrently contribute to the difficulty faced by CI users in spatial hearing. Poor temporal resolution typically experienced by CI users[11], prevents them from taking advantage of interaural time difference (ITD) cues [12]. In addition, CI users are also less sensitive to level changes [13], implying that they could not benefit as much as NH listeners from interaural level difference (ILD) cues for sound localization.

In the last few years, some studies [2], [5], [6], [14]–[16] have examined the different factors that affect spatial selective auditory attention with hearing impaired listeners. One way of examining spatial hearing with headphones is to determine from where a sound is coming which is known as lateralization. Strelcyk and Dau [17] investigated the effect of stimuli level on sound lateralization for NH and HI listeners. They measured the lateralization threshold for 750 Hz tones, fixed at 70 and 35 dB SPL. The results showed that generally, the lateralization threshold improved at the higher stimulus level comparing to the lower stimulus level. The difference between NH and HI listeners' performance at the lower tone level was smaller than that at the higher tone level. Consistent with this result, Smoski [18] observed a smaller deficit of HI listeners' lateralization (relative to NH) at a lower tone level. However, HI listeners in the study by Hawkins and Wightman [19] showed a smaller lateralization deficit at a higher stimulus level than that at a lower stimulus level, when the narrow band noise was used as stimulus in a quiet condition. The different outcome of these studies [18], [19] may be attributed to the type of stimuli (tone/noise) used in these studies. Strelcyk and Dau explained that at higher stimulus levels, lateralization judgement could arise from the excitation of a larger area of the basilar membrane instead of the local excitation area [17]. NH listeners may take advantage of the excitation spread particularly towards places corresponding to high frequencies and integrate the additional information placed at high frequency areas. However, HI listeners with high frequency sensory neural hearing loss may not benefit from this additional information, because it falls in the sloping region of their hearing loss. Nevertheless, the excitation model may not be directly applicable to the cochlear implant participants' performance.

The spatial separation between the target speech and competing sounds also plays an important role in spatial selective auditory attention. Spatial separation helps listeners to segregate the sound sources and consequently improves the speech intelligibility. The difference between speech reception threshold (SRT) in co-located and spatially separated sound sources, which is known as spatial release from masking (SRM) [20], is commonly used to measure the benefits of spatial separation. Gallun et al. [21] compared the SRTs of HI listeners when target sentence levels were presented at 19.5 dB SL and 39.5 dB SL in quiet. In an adaptive approach, the levels of the maskers were adjusted relative to the level of the target sentences to estimate the masked threshold [target-to-masker ratio (TMR) giving 50% correct]. The results showed that the masked threshold and SRM improved with an increase in SL indicating that the listener's spatial auditory selective attention covaries with audibility.

Studies above have collectively shown evidences that the level of sound do affect the spatial auditory selective attention in both normal hearing and hearing impaired listeners. To best of our knowledge, there is little existing work but a growing effort in examining the effect of sound level on spatial auditory selective attention in CI users.

Our team [22] has previously shown the effect of speech level on the speech quality as perceived by normal hearing listeners and CI users. The study showed that even at the same signal-to-noise ratio (SNR), but at different conversational speech levels, different patterns of perceived quality judgement were observed in NH listeners and CI users. While the NH listeners preferred higher speech levels, CI listeners preferred lower speech levels, suggesting that CI listeners rather choose lower noise levels at the expense of poorer speech audibility. Such evidence of perceptual difference between two groups inspired us to examine the effect of speech level on the spatial auditory selective attention with NH and CI listeners.

Spatial hearing is commonly assessed from behavioral responses, but it can also be observed in the electrophysiological response. Some studies has also shown that spectral and temporal features of attended speech can be extracted from the cortical response [23], [24]. In [25], electroencephalography (EEG) signal was recorded on the scalp of NH listeners while they engaged in a two speakers cocktail party scenario. The listeners were instructed to attend one of the speakers and ignore the other one. A linear decoder was trained to reconstruct the speech envelope using the EEG signal. Listeners' locus of attention was determined reliably based on the speech envelope reconstructed from the EEG signal. It was shown in [26] that the spearman correlation between original speech envelope and the speech envelope reconstructed from the EEG signal increased by increasing the SNR. They demonstrated that the behaviorally measured speech intelligibility was highly correlated with the congruence between the original and 'reconstructed' envelope. Such electrophysiological EEG approach was thought to be another applicable measure that drive us to better understanding of spatial hearing for NH and CI listeners in this study.

In current study, we examined the effect of stimulus levels on the spatial selective auditory attention in the speech-on-speech masking for NH and CI listeners. Target and maskers stimuli were presented at different conversational levels, while keeping TMR the same to examine the effect of the speech level independent of the effect of the TMR. In addition to behavioral experiments, the accuracy of the attended speech envelope reconstruction from the EEG signals were computed as metric to indirectly measure the auditory attention from cortical response in three speakers cocktail party problem. The main purpose of this study is to examine the effect of speech levels on spatial hearing ability, for NH and CI listeners, in both behavioral and neural outcomes.

## 2- Method

### 2-1- Participants

Eight NH listeners (four male; mean age: 24, range: 21-30, SD: 3.5 years) and nine CI users (five male; mean age: 58, range: 24-77, SD: 22 years) participated in this study. All participants self-reported with no history of cognitive deficits prior to participation. All NH subjects were verified with 20 dB HL or lower across all octave frequencies between 250 to 8000 Hz in their pure tone audiograms. Unaided hearing thresholds were identified for CI group over the same octave frequency range. The demographic information for CI users is shown in Table 1. All procedures were approved by institutional review board of University of Texas at Dallas.

*Table 1- Demographic information of CI listeners*

| Subject number | Age (years) | Gender | CI Ear | CI Model | CI use (years) | Speech processing strategy | Duration of HL (years) | Pure tone average of 500, 1000, and 2000 Hz (dB HL) | Etiology |
|---|---|---|---|---|---|---|---|---|---|
| Bilateral CI | | | | | | | | | |
| 1 | 75 | Male | Both | Medel/ Sonnet | 7 | FS4 | 33 | Right: NR Left: NR | Meniere's Disease |
| 2 | 40 | Female | Both | Medel/ Sonnet | 4 | FS4 | 38.5 | Right: NR Left: NR | unknown |
| 3 | 68 | Male | Both | Medel/ Sonnet | 10 | FS4 | 38 | Right: NR Left: NR | unknown |
| 4 | 25 | Female | Both | Cochlear/ Nucleus 6 | 21 | ACE | 25 | Right: NR Left: NR | unknown |

| 5* | 77 | Female | Both | Medel/ Sonnet | 70 | FS4 | 6 | Right: NR Left: NR | meningitis |
|---|---|---|---|---|---|---|---|---|---|
| Unilateral CI | | | | | | | | | |
| 6 | 67 | Male | Right | Cochlear/ Nucleus 6 | 11 | ACE | 11 | Right: NR Left: 70 | Noise induced |
| 7 | 67 | Male | Right | Medel/ Sonnet | 10 | FS4 | 24 | Right: NR Left: 71.6 | Meniere's Disease |
| 8* | 24 | Female | Right | Cochlear/ Nucleus 7 | 1 | ACE | 23 | Right: NR Left: NR | Genetic |
| Bimodal hearing | | | | | | | | | |
| 9 | 77 | Male | Right | Medel/ Sonnet | 2 | FS4 | 20 | Right: 82.5 Left: 66.6 | Noise induced |

* Subject 5 and 8 from CI listeners and two of the NH listeners were not able to participate in EEG recording

## 2-2- Behavioral experiment

### 2-2-1- Stimuli

We adopted a matrix sentence test [27] for the behavioral spatial hearing test. On each trial, three spatially separated sentences (one target and two maskers) were presented to the subjects via different loud speakers. The sentences were formed by concatenating the words, with one word from each of the five categories of words (name, verb, number, adjective, and noun). Each category consists of 8 words that were repeatedly spoken by 18 female and 18 male talkers. Table 2 shows the eight words in each of the five categories. The sentences are grammatically correct and sound natural, but conceptually unpredictable to minimize the effect of higher order language processing.

*Table 2- Matrix sentences*

| **Name** | **Verb** | **Number** | **Adjective** | **Noun** |
|---|---|---|---|---|
| Jane | Took | Two | New | Toys |
| Gene | Gave | Three | Old | Hats |
| Pat | Lost | Four | Big | Shoes |
| Bob | Found | Five | Small | Cards |
| Sue | Bought | Six | Red | Pens |
| Mike | Sold | Seven | Blue | Socks |

| Lynn | Held | Eight | Cold | Bags |
|------|------|-------|------|------|
| Jill | Saw  | Nine  | Hot  | Gloves |

*2-2-2- Procedure*

The experiments were conducted in a double-wall soundproof booth. Five speakers placed at -90°, -45°, 0°, 45°, and 90° azimuth were used to present the stimuli. The speakers located at the radius of 1 meter away from subjects at the height of the subjects' head. A touchscreen monitor was located in front of the subject.

In each trial, three different talkers were randomly selected to present the target and maskers sentences. Three different sentences were selected from the words listed in Table 2, one as a target and two as maskers. Speech recognition performance for five speaker configurations were examined (Figure 1). To explore the effect of speech level on spatial selective auditory attention, three different levels at fixed TMR were examined as follows. For CI listeners stimuli were presented at 10 dB TMR, with target level at 75, 65, and 55 dB SPL and maskers level at 65, 55, and 45 dB SPL, respectively. For NH listeners, to avoid ceiling effect, stimuli were presented at 0 dB TMR, with target level at 75, 65, and 55 dB SPL and maskers level at 75, 65, and 55 dB SPL.

At the beginning of each trial, an arrow appeared on the monitor indicating the location of the target speaker that subject should attend to. After the subject pressed a start button, stimuli were presented and waited infinitely long until subject complete the answer. Once speech stimuli were presented, subject should select the words uttered by the target speaker in the table similar to Table 2 appeared on the monitor. Subjects were instructed to guess a word if they were not able to identify the word.

The experiment was conducted in three sessions associated with three different stimuli levels, which stretches over one day visit. At each session, 50 trials (five target speaker configurations * 10 repetitions) were randomly presented to the subject. As mentioned earlier, sentences, words, and talkers were selected randomly for each trial. Before the main experiments, each subject participated in one training session to get familiar with the procedure. At the training session, target stimuli were presented at 65 dB SPL. The maskers were presented at 55 dB SPL for CI listeners and 65 dB SPL for NH listeners.

For bilateral CI user, the experiment was repeated three times; one with both CIs turned on, one with only left CI turned on , and one with only right CI turned on. For unilateral CI users who had residual hearing in contralateral ear, the same experiment was repeated twice; one with CI turned on, and one with CI turned off. For bimodal CI listener (with hearing aid (HA) on the contralateral ear), the experiment was again repeated three times; one with both CI and HA turned on, one with only CI turned on and HA turned off , and one with only CI turned off and HA turned on.

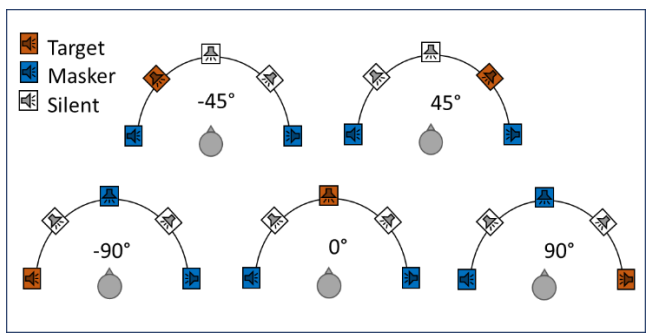

*Figure 1- Speakers configuration*

## 2-3- Electrophysiological experiment

### 2-3-1- EEG Recording setup

The BrainVision system (actiCHamp amplifier) was used to obtain EEG signals via a 64-channel actiCAP Electrode Cap. Scalp electrode placement was set in accordance with international 10-20 system. The ground electrode was placed at FPz, and the reference electrode was placed at FCz of subject's head. Horizontal and vertical ocular artifacts were recorded using additional electrodes connected to two bipolar adaptors (BIP2AUX). All electrodes were kept at impedance lower than 10 kΩ. EEG signal was digitally recorded using a sampling rate of 1 kHz. The data were stored for offline analysis. Participants were seat in the middle of sound booth wearing scalp electrode cap on their head. They were asked to keep calm, look at a fixed point on the monitor, and minimize eye blinking and muscle movement while sound was presented.

This study attempted to extract the cortical activities that are entrained to the speech envelope using a decoder. Our EEG experiment consisted of two recording sessions. One session was for collecting data on the speech in quiet to train the decoder which extracts the speech envelope from EEG signal. The other session was for measuring EEG associated with different spatial configuration and sound level conditions in a three speakers cocktail party scenario.

### 2-3-2- Training the decoder

First, EEG was recorded with a continuous speech passage and used as the data to train the decoder. The subjects listened to two short story narrations. The first story was "lady or tiger" narrated by a female speaker (duration = 632.7341 sec), and the second story was "ambitious guest" narrated by a male speaker (duration 615.2858 sec). Both speakers were native American English speakers. Silent gaps (more than 300msec in duration) were removed from the recordings. The stories presented at 65 dB SPL from front speaker without any noise. No question was asked before and after presenting these stimuli. EEG data was recording while the subject was listening to the stimuli.

### 2-3-3- Test in the three speakers cocktail party scenario

We presented one target and two maskers to emulate a cocktail party scenario. The stimuli were short passages with the duration of 38 to 45 seconds. Target and maskers passages were selected randomly from Connected Speech Test (CST) [28] and Speech Intelligibility Rating (SIR) [29] data set. To remove the effect of the speaker gender, half of the trials had a female target speaker and the other half had a male target speaker. For the trials with female target, maskers were male, and for the trials with male target, maskers were female. Three speaker configurations were examined for different permutation of target and maskers location using the speakers at -90, 0, and 90 azimuth. The stimuli were presented at three different levels with fixed TMR at 10 dB. Targets were presented at 75, 65, and 55 dB SPL and maskers at 65, 55, and 45 dB SPL, respectively. A total of 18 combinations (3 speaker configurations*2 target genders*3 speech levels) were randomly presented to each subject as separate trials. At the beginning of each trial an arrow was appeared on the monitor showing the location of the target speech that subject should attend to. After a one second pause, the stimuli were presented. After finishing the presentation of each trial, subjects should respond to two questions about target passage. The aim of these questions was to keep the subject's attention to the target speech.

### *2-3-4- Detecting the auditory attention from EEG signal*

To detect the auditory attention in the cocktail party scenario, we reconstructed the attended speech envelope from EEG signal. We needed a decoder that maps the speech envelope to the corresponding EEG signal. Individualized decoder for each subject was trained using the presented speech and recorded EEG signal in training session. We used mTRF toolbox (version 1.5) [30] to train the decoder and reconstruct the attended speech envelope from neural response.

The speech envelopes of the stimuli were extracted using Hilbert transform, and band pass filtered to 0.3-30 Hz using filtfilt function. To decrease the processing time, speech envelopes were down sampled to 128 Hz and normalized to a range of value between 0.0 and 1.0 .

After preprocessing of the EEG signal recorded from 64 channels and removing the artifacts using independent component analysis (ICA), EEG waves were band pass filtered to 0.3-30 Hz using Fieldtrip toolbox [31]. The filtered EEG were down sampled to 128 Hz and normalized to [0-1]. All the implementations were in MATLAB 2018b.

Suppose that the linear mapping from the instantaneous neural response $r(t,n)$, sampled at times $t = 1, \ldots, T$ and at channel $n$, to the speech envelope $s(t)$ is represented as $d(\tau, n)$ which is the decoder that integrates the neural response over a range of time lags $\tau$. The linear convolution equation corresponding to this system could be expressed as:

$$\hat{s}(t) = \sum_n \sum_\tau r(t+\tau, n) d(\tau, n), \qquad (1)$$

In equation 1, the $\hat{s}(t)$ is the reconstructed speech envelope. The cost function to design the decoder is the MSE between $s(t)$ and $\hat{s}(t)$.

$$\min \varepsilon(t) = \sum_t [s(t) - \hat{s}(t)]^2, \qquad (2)$$

Which results in the following formula for decoder:

$$d = (R^T R)^{-1} R^T s, \qquad (3)$$

In the above equation, $R$ represents the lagged time series of neural response matrix $r$. If $R_c$ contains all time lags $[\tau_{min}, \tau_{max}]$ for channel $c$, then:

$$R = [R_1\ R_2\ \cdots\ R_N], \qquad (4)$$

Where:

$$R_c = \begin{bmatrix} r(1-\tau_{min}, c) & r(-\tau_{min}, c) & \cdots & r(1, c) & 0 & \cdots & 0 \\ \vdots & \vdots & \cdots & \vdots & r(1, c) & \cdots & \vdots \\ \vdots & \vdots & \cdots & \vdots & \vdots & \cdots & 0 \\ \vdots & \vdots & \cdots & \vdots & \vdots & \cdots & r(1, c) \\ r(T, c) & \vdots & \cdots & \vdots & \vdots & \cdots & \vdots \\ 0 & r(t, c) & \cdots & \vdots & \vdots & \cdots & \vdots \\ \vdots & 0 & \cdots & \vdots & \vdots & \cdots & \vdots \\ \vdots & \vdots & \cdots & \vdots & \vdots & \cdots & \vdots \\ 0 & 0 & \cdots & r(T, c) & r(T-1, c) & \cdots & r(T-\tau_{max}, c) \end{bmatrix}, \qquad (5)$$

To avoid overfitting of the model to the training data, Tikhonov regularization is applied on the optimization problem of equation 2, which results in:

$$d = (R^T R + \lambda I)^{-1} R^T s, \qquad (6)$$

In the above equation, $\lambda$ is the regularization parameter.

To find the best $\lambda$, we used leave one out cross validation. The training data (speech stimuli and corresponding EEG signal collected in training session) was split to $K$ trials. The preliminary decoders constructed from single trials would be $\tilde{d}_k = (R_{(k)}^T R_{(k)} + \lambda I)^{-1} R_{(k)}^T S_{(k)}$, where $R_{(k)}$ and $S_{(k)}$ are the EEG response and stimuli of trial $k$ (see equation 6). The preliminary decoders from all trials except trial $k$ were averaged to define decoder $d_k$ which decodes trial $k$.

$$d_k = \frac{1}{K-1} \sum_{\substack{i=1 \\ i \neq k}}^{K} \tilde{d}_i \qquad k = 1, \cdots, K \tag{7}$$

Using the decoders defined by equation 7 for each trial, the speech envelopes were reconstructed. The Pearson correlation between the reconstructed speech envelope and the original speech envelope is a criteria to evaluate the accuracy of the decoder. The $\lambda$ that resulted in maximum correlation averaged across $K$ trials was selected to train the decoder. Here, we split the training data of each subject to 20 trials and examined a range of $\lambda = \{10^{-3}, 10^{-2}, \cdots, 10^{11}\}$.

Different CI users receive different signals from their CIs, based on the processing strategies and specific artifacts caused by different RF transmission systems. Therefore, we selected individualized $\lambda$ value for each CI user using each individual's training data. For NH listeners, we selected a common $\lambda$ value of $10^5$. The $\lambda$ for each CI subject is shown in **Error! Reference source not found.**.

*Table 3- Selected $\lambda$ for CI subjects*

| Subject# | CI-1 | CI-2 | CI-3 | CI-4 | CI-6 | CI-7 | CI-9 |
|---|---|---|---|---|---|---|---|
| $\lambda$ | $10^1$ | $10^5$ | $10^1$ | $10^5$ | $10^{-1}$ | $10^3$ | $10^1$ |

### *2-3-5- EEG based speech detection accuracy*

Having the individualized linear decoder designed for each subject, the speech envelope of the attended speech in the three speaker cocktail party test was reconstructed from EEG signal. Pearson correlation between reconstructed speech envelope $\hat{s}(t)$ and each stimulus envelope, target $s_T(t)$ and two maskers $s_{M1}(t)$, and $s_{M2}(t)$, was calculated and referred to as $r_T$, $r_{M1}$, and $r_{M2}$, respectively. The stimulus having higher correlation with reconstructed speech envelope was detected as attended speech. For example, if $r_T$ was greater than $r_{M1}$ and $r_{M2}$, the target speech was detected as attended speech. The number of the trials in which the target stimulus was detected as attended speech over the total number of trials is referred to as EEG-based speech detection accuracy. Figure 2 summarizes this procedure.

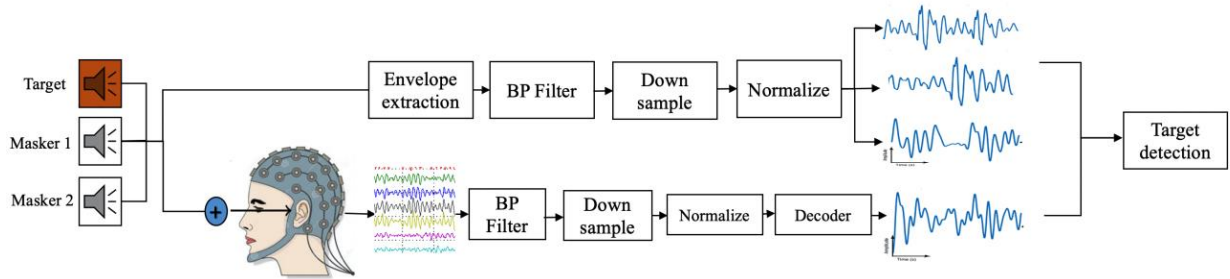

*Figure 2- Procedure of the attended speech detection from EEG signal in three speakers cocktail party test*

## 3- Results

### 3-1- Behavioral speech recognition score

As one of our goals was to investigate the spatial selective auditory attention in electrical and acoustic hearing, we grouped our CI data into three hearing categories: 1- Electrical hearing only (Bilateral CIs + unilateral CI with no residual hearing in contralateral ear + bimodal CI when the HA is off), 2- Acoustic hearing only (Unilateral CIs with residual hearing in contralateral ear when CI is off + bimodal CI when CI is off), and 3- Electric+ acoustic hearing (Unilateral CIs with residual hearing in contralateral ear + bimodal CI). From now, we use EH for the group of electrical hearing only, AH for the group of acoustic hearing only, and EAH for the group of electric+acoustic hearing. Among nine CI subjects, six subjects were in EH group (five bilateral CI users, one unilateral CI user who had CI in right ear and no residual hearing in left ear), but three subjects were in EH and EAH group (two unilateral CIs and one bimodal CI).

The speech recognition scores were calculated as the number of correct words chosen by subjects over the total number of words. The scores are represented in percentage for EH, AH and EAH CI users in Figure 3-a, -b and -c, respectively. In EH category, the speech recognition score decreased as the target sound level increased in a constant TMR. In contrast, in AH category, increasing the target speech level resulted in increased speech recognition score. In EAH category, the speech recognition score decreased by increasing the target level from 55 dB SPL to 65 dB SPL, and increased by increasing the target level from 65 dB SPL to 75 dB SPL. A one way analysis of variance (ANOVA) was conducted on speech recognition score with factor of target level for each group. The results showed that there was a significant main effect of target level in EH ($F(2,2547)=18.22$, $p<0.01$), AH ($F(2,447)=9.04$, $p<0.01$), and EAH ($F(2,447)=3.8$, $p=0.02$). A pairwise comparison with Bonferroni adjustment showed that speech recognition score for EH category was significantly higher when the target level was at 55 dB SPL than that when the target level was at 65 and 75 dB SPL, and speech recognition score was significantly higher when the target level was at 65 dB SPL than that when the target level was at 75 dB SPL. The pairwise comparison for speech recognition score in AH category showed that the speech recognition score increased by increasing the target level. The speech recognition score at 55 dB SPL target was significantly lower than the speech recognition score at 65 and 75 dB SPL, but there was no significant difference between speech recognition score when the target level was 65 and 75 dB SPL. For EAH category, speech recognition score at 65 dB SPL target was significantly lower than that at 75 dB SPL target. The asterisk(*) in Figure 3 shows the significant difference between the speech recognition score at different target levels. The speech recognition scores for NH group were nearly equal across all three speech levels (Figure 3-d). The ANOVA test showed no significant effect of target speech level on the speech recognition score for NH subjects.

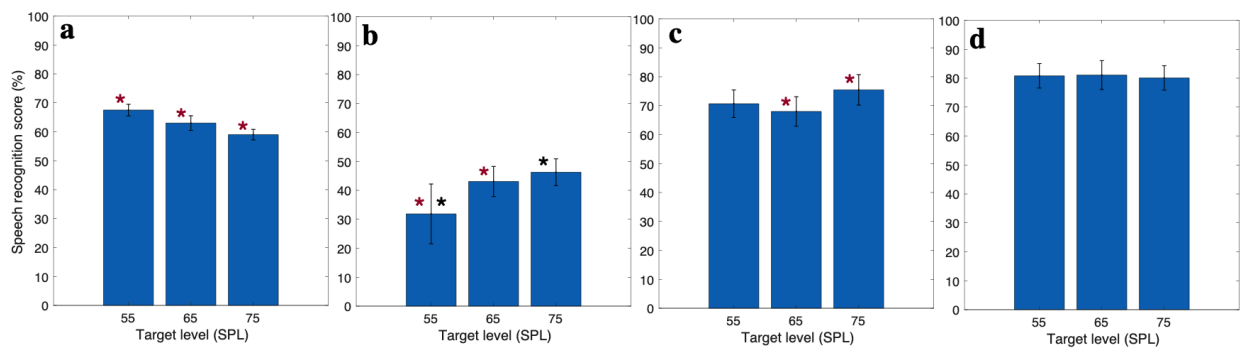

*Figure 3-Speech recognition score for a) EH, b) AH, c) EAH CI listeners, and d) NH listeners.*

## 3-2- EEG based speech detection accuracy

With the individualized decoder for each subject, the EEG data recorded in the three speakers cocktail party experiment was decoded to find the attended speech, as explained in section 2-3. The speech detection accuracy for different target speech levels is presented in Figure 4-a, for CI subjects in EH category, Figure

4-b for AH category, and Figure 4-c for EAH category. The statistical analysis ANOVA showed a significant effect of target level on the target detection accuracy in EH category. As Figure 4-a depicts, by increasing the target level (while keeping the TMR at 10), the speech detection accuracy decreased. The pairwise comparison with Bonferroni adjustment for EH category showed that the speech detection accuracy at 55 dB SPL target was significantly lower than that at 75 dB SPL target. The asterisk (*) in Figure 4 presents the significant difference in speech detection accuracy between different target levels. There was no effect of target level on speech detection accuracy in AH and EAH categories. The speech detection accuracy for different target levels is presented in Figure 4-d for NH subjects. The results show decreasing pattern of speech detection accuracy as the target level increased. Despite the pattern, the ANOVA showed no significant effect of target level on the speech detection accuracy for NH subjects.

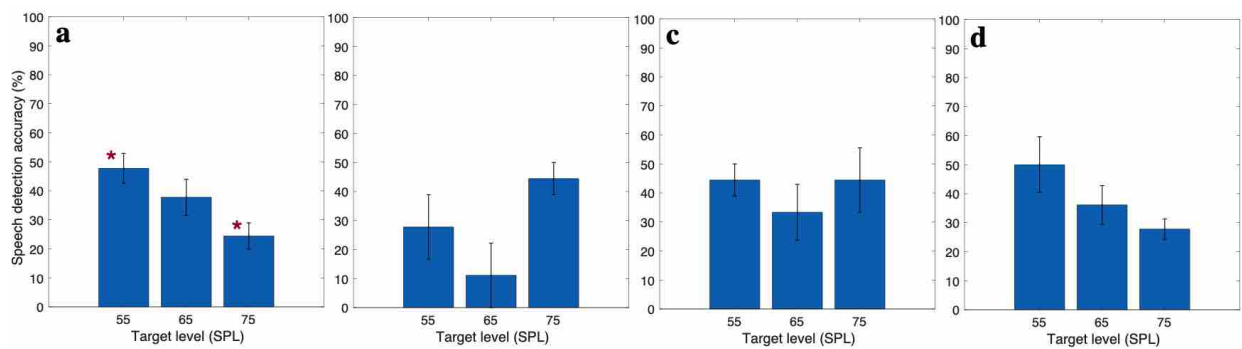

*Figure 4- Speech detection accuracy for CI subjects in case of a) EH, b) AH, and c) EAH CI listeners, and d) NH listeners.*

## 3-3- Case report for individual subject

Considering significant variability in hearing modality and background among our CI subjects and small sample size, we are presenting the analysis outcome on 'case-by-case' basis. Each CI subject's level-dependent spatial hearing patterns were individually demonstrated associated with their hearing characteristics. Subject CI 5 and CI 8 have not completed the physiological part of the experiment and are not included in the analysis.

### *3-3-1- Subject CI 1*

CI 1 (75 years old male) is a bilateral CI user with no residual hearing in both ears. He showed better speech recognition scores with bilateral hearing compared to those with unilateral hearing with one of his CIs off. His scores were likely to be higher at lower target speech level (55 dB SPL) compared to higher target speech level (75 dB SPL). His spatial hearing scores were higher when speech was presented at the direction where his CI is on.

Neural outcomes represented that higher neural tracking accuracy was associated with lower conversational level of speech in both unilateral CI conditions. Speech detection of neural tracking was the highest when target speech was presented from 0° azimuth. This neural detection accuracy pattern is somewhat discrepant with his behavioral outcome where evidences of directional benefits of CI is provided.

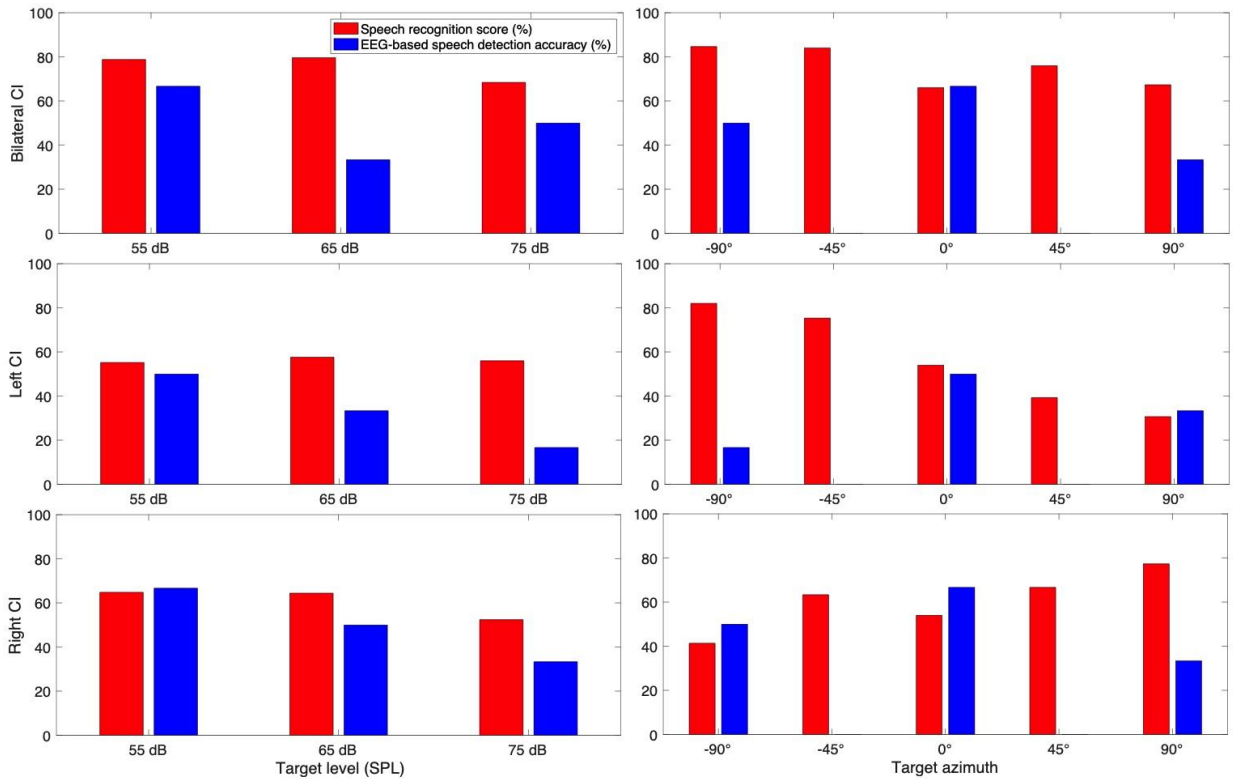

*Figure 5- Speech recognition score and speech detection accuracy from EEG for CI1.*

### 3-3-2- Subject CI 2

CI 2 (40 years old female) is a prelingually deafened CI user in both ears. Her speech recognition scores were nearly equal across the levels for bilateral condition. On the other hand, scores for 75 dB SPL condition was significantly lower than scores for 55 and 65 dB SPL conditions when only one CI was used. A consistent trend of directional benefits were observed when either left or right CI was on.

In her overall neural outcomes, clear pattern of level and directionality was not found.

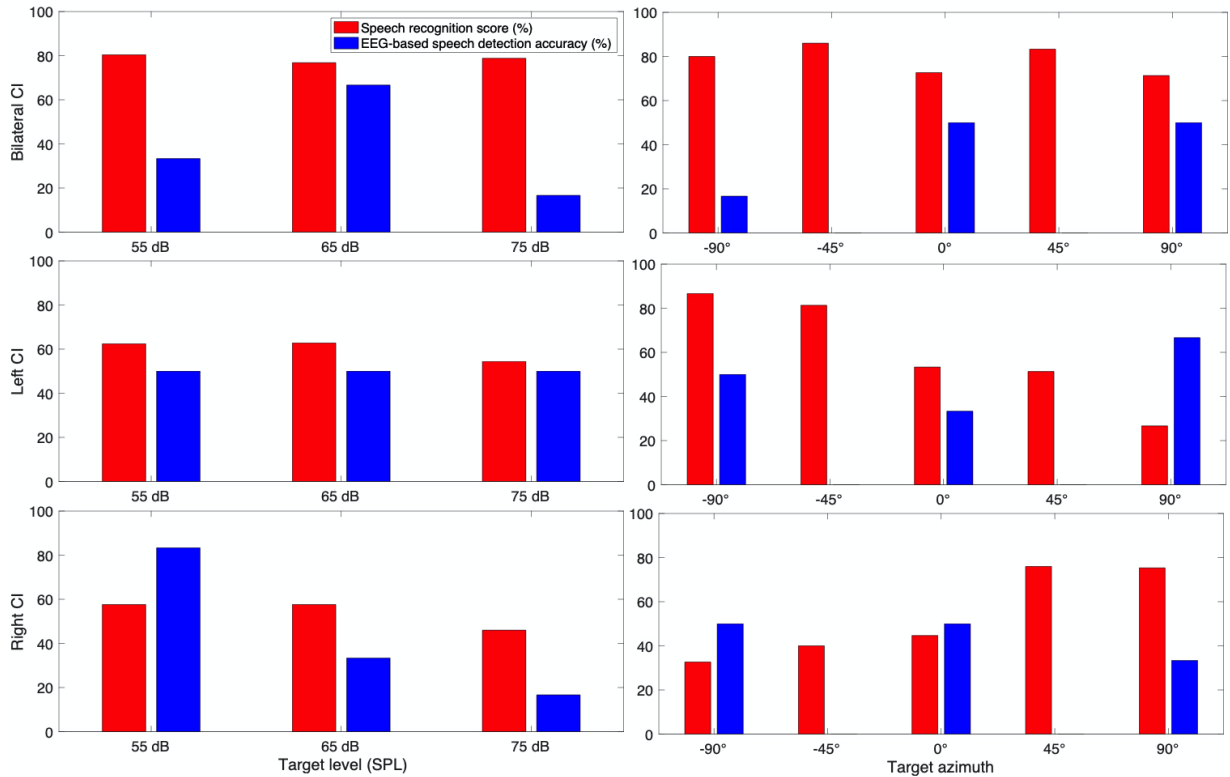

Figure 6- Speech recognition score and speech detection accuracy from EEG for CI2.

### 3-3-3- Subject CI 3

CI 3 (68 years old male) is a post lingually deafened bilateral CI. In behavioral speech recognition test, he showed the advantage of bilateral hearing over unilateral hearing. There was a level effect in which slightly higher scores were shown for lower target speech level compared to higher target speech level. Clear trend of directional benefit of CI is also represented.

In neural detection accuracy, higher EEG-based speech detection accuracies with lower speech levels and closer directions to the target speech azimuth were observed despite few exceptions including the left CI condition showing better neural accuracy at 90°(right), and the right CI condition showing 0 point of neural accuracy at 65 dB SPL.

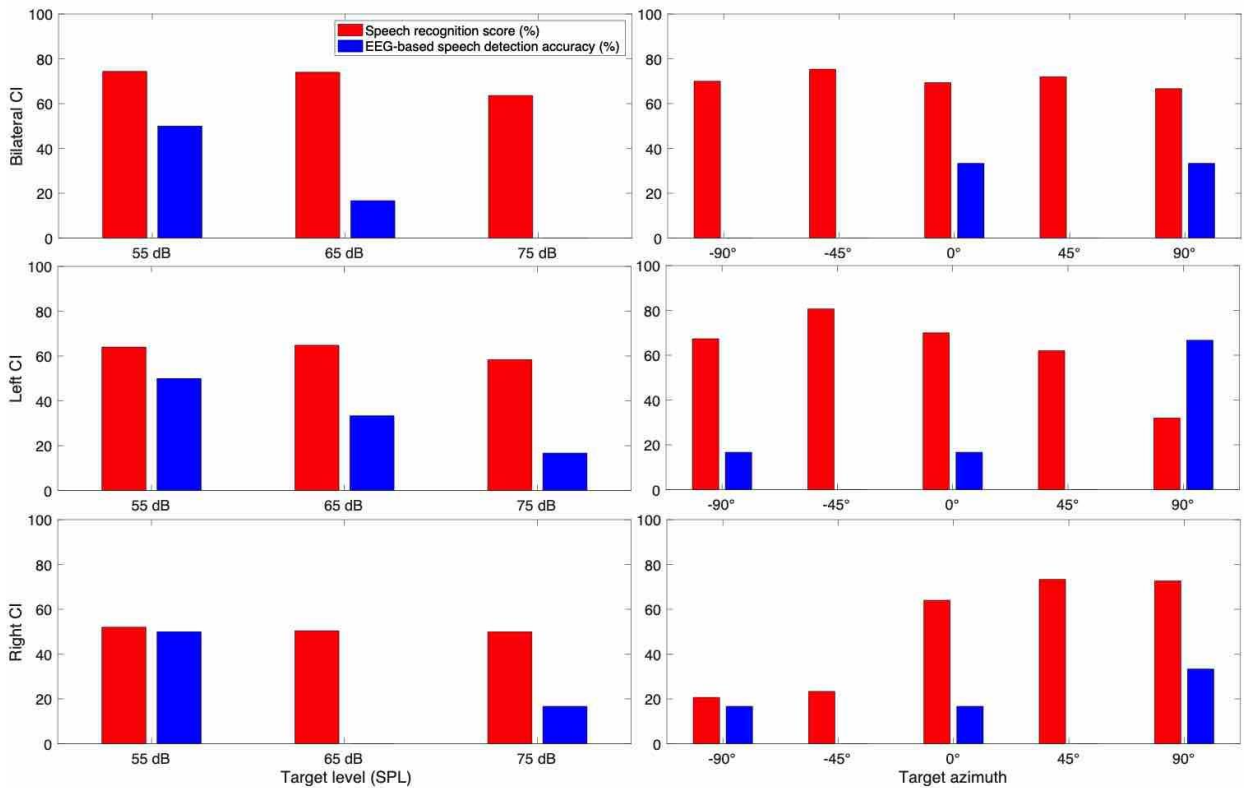

*Figure 7- Speech recognition score and speech detection accuracy from EEG for CI3.*

### 3-3-4- Subject CI 4

CI 4 (25 years old female) showed the advantage of bilateral hearing over the unilateral hearing at all three sound levels. By increasing the sound level, the speech recognition score decreased. The spatial speech recognition score was higher at the azimuths closer to direction of activated CI.

EEG results showed that except for bilateral condition, speech detection accuracy decreased as the target level increased. The pattern of speech detection accuracy based on different target azimuth is not consistent with behavioral results. We expected to have higher speech detection accuracy at the azimuths closer to amplification side, but the results show a decreasing pattern of speech detection accuracy as the target goes from 90˚ to the -90˚, in bilateral and left CI condition. The speech detection accuracy is less than chance level (33%) in right CI condition at all target azimuths.

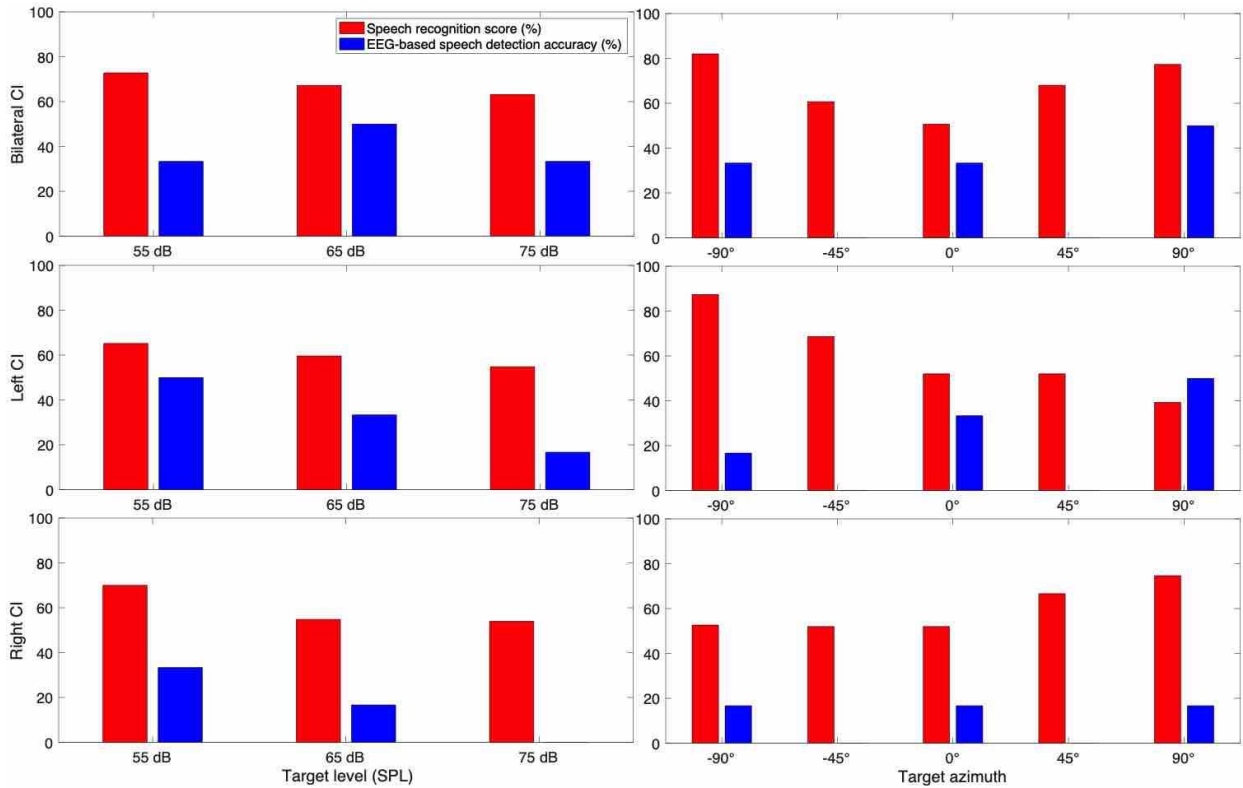

*Figure 8- Speech recognition score and speech detection accuracy from EEG for CI4.*

### 3-3-5- Subject CI 6

CI 6 (67 years old male) was a unilateral (right) CI user having 70 dB HL of pure tone average (PTA) in his contralateral ear. The PTA is calculated by averaging the hearing threshold at 500 Hz, 1 kHz, and 2 kHz frequencies. He receives significant amount of electrical energy via his right ear CI and receives some amount of acoustic energy throughout his left ear in either CI is on or off condition. Unlike EH group (e.g., CI 1, CI 2, CI 3, and CI 4), speech recognition scores for this subject increased by increasing the speech level in both CI-on and CI-off conditions which shows taking advantage of residual hearing in higher sound levels. His spatial hearing outcomes in CI-on condition clearly indicated CI benefits in right ear showing higher scores at the conditions where speeches were presented from azimuths closer to the CI side.

The level effect shown in the results of EEG test was different from that of behavioral test. Spatial hearing outcomes showed some indications of CI benefits on right ear in CI-on condition, but they were not very systematic in CI-off condition. In CI-off condition (where he relies only on left ear), we expected to have higher speech detection accuracy at 0° compared to 90°. However in Figure 9, the speech detection accuracy at 90° is higher than that at 0°.

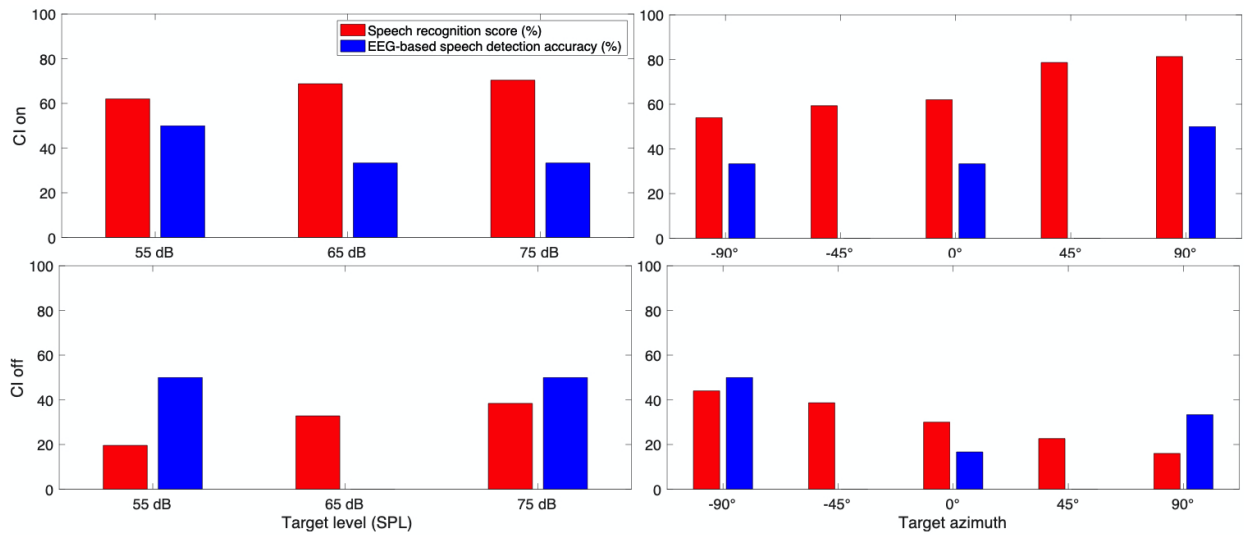

*Figure 9- Speech recognition score and speech detection accuracy from EEG for CI6.*

### 3-3-6- Subject CI 7

CI 7 (67 years old male) was a unilateral (right) CI user with 71.6 dB HL of PTA in his contralateral ear. Thus, both CI 6 and CI 7 are similar in terms of their degree of residual hearing and hearing modality. His speech recognition score was higher at higher presentation level of 75 dB in CI-on condition as well as CI-off condition. Consistent with other individuals' spatial hearing pattern, his spatial hearing tends to be better at the direction that provides him with a better audibility.

Overall, neural tracking of CI 7 appeared to be consistent with his behavioral pattern when his CI was on. Few exceptions, however, were found in CI-off condition (EEG-based speech detection accuracy for 65 dB SPL was lower than that for 55 dB SPL, and at -90° was lower than that at either 0° or 90°).

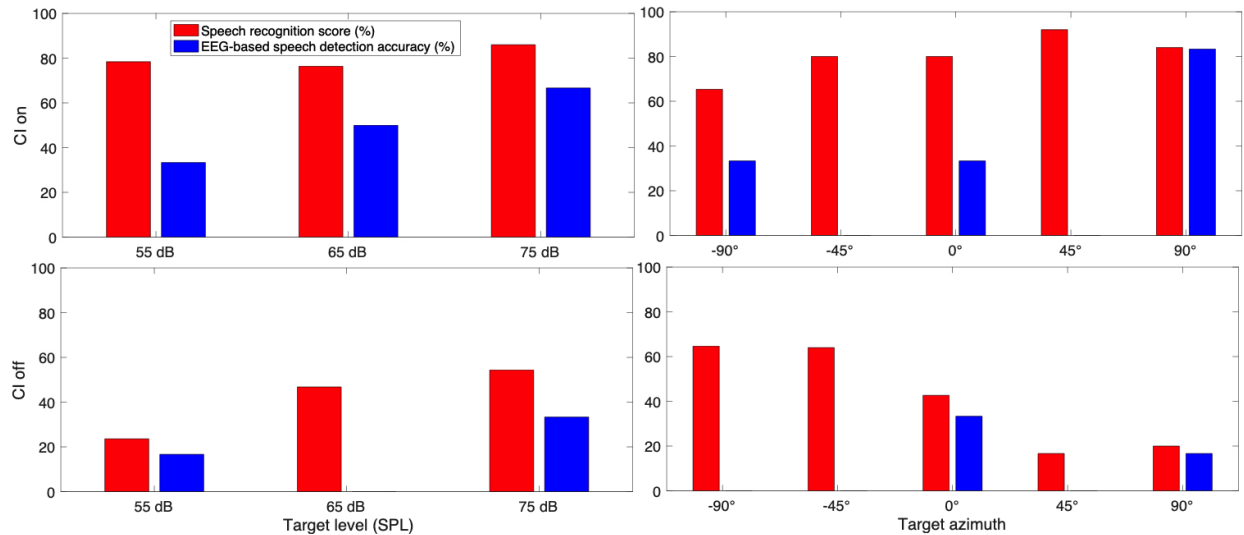

*Figure 10- Speech recognition score and speech detection accuracy from EEG for CI7.*

### 3-3-7- Subject CI 9

CI 9 (77 years old male) was the sole bimodal subject of our study who use CI in his right ear and HA in his left ear. This type of subject is known to receive a great benefit from his acoustic hearing via HA as well as CI. Our behavioral result showed that his perceptual performance was higher when using CI compared to HA. Bimodal benefits was shown in 55 dB SPL and 75 dB SPL condition, but not in 65 dB SPL condition. When CI was on, higher score was obtained at the softer target speech level, but this effect disappeared with the addition of acoustic hearing. Spatial hearing performances of him followed the general trend showing higher scores at the direction his amplification was placed.

EEG outcome was not consistent with behavioral outcomes showing some nonsystematic patterns as a function of speech level and target direction. In contrast to the behavioral result, in HA-on condition, the speech recognition score increased by increasing the sound level. There is a clear directional pattern in CI-on condition in which higher EEG-based speech detection accuracy was acquired at the azimuths closer to the hearing side.

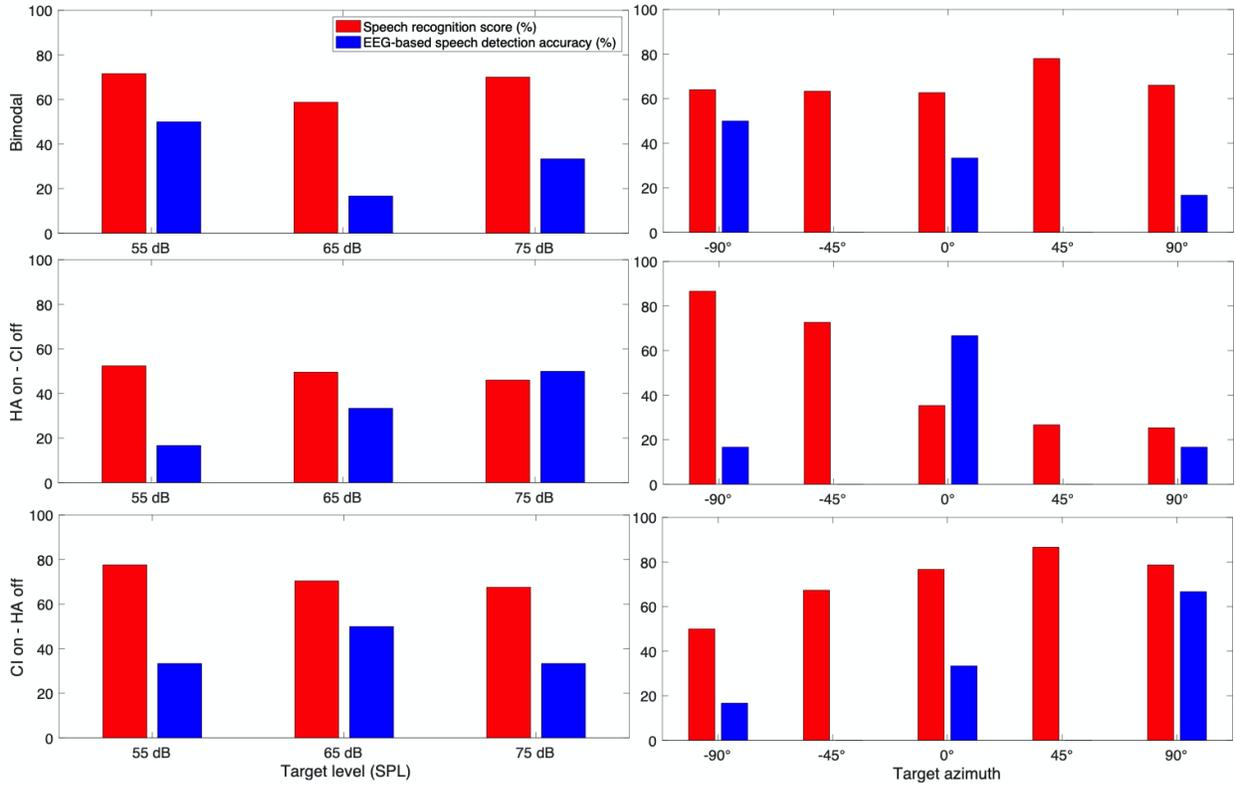

*Figure 11- Speech recognition score and speech detection accuracy from EEG for CI9.*

## 4- Discussion

Our behavioral test results presented no statistical difference in NHs' speech recognition score under different conversational sound levels, however we found that sound level significantly affected CIs' speech recognition score. The electrical dynamic range of CIs differs from the dynamic range of NH, thus the loudness that CI listeners perceive may be different from that of which NH listeners perceive. It is reasonable to assume that input dynamic range for NH would approximately be 120 dB, considering the loudness comfort NH listeners generally show. However, CIs substantially compress the acoustic dynamic range into the much narrower electrical dynamic range in the system. The electrical dynamic range varies from 10 to 80 dB depending on the signal processing strategy and CI manufactures [32]. We infer that the compression applied at higher levels may distort the speech recognition in CI users which negatively affects their spatial selective auditory attention.

In the case of electrical hearing only (EH), CI listeners speech recognition decreased by increasing the sound level. Some studies showed that in a quiet condition, CI listeners' speech recognition improves by increasing the speech level [33]. However, our previous study showed that increasing the speech level in the presence of noise at constant SNR degrades the perceived quality of speech by CI listeners [34]. It is generally reported that CI users are more susceptible to noise than NH listeners when recognizing speech. We conclude that in spatial hearing, it is more difficult for CI listeners to suppress the higher masker level accompanied with the increased target level, which resulted in reduced spatial selective auditory attention in the CI group. The CI listeners data in acoustic hearing only (EH) category (unilateral CI users with residual hearing in their non-implanted ear and bimodal CI user, when their CI was off), showed an improvement in their spatial selective auditory attention with increase on the sound level. It can be speculated that the negative effect of increased masker levels is less than the positive effect of increased target levels in less audible listening situation. The electrical hearing seems more vulnerable to the amount of noise even at the same TMR. In other words, the negative effect of increased maskers' level is perceptually greater than the advantage of the higher audibility of target. In the case of electric+acoustic hearing (EAH), the results show that the spatial selective auditory attention with target level at 55 dB SPL relies more on electrical hearing (see Figure 3-c). By increasing the target level to 65 dB SPL, electrical hearing degraded while the improvement in the acoustic hearing was not enough to compensate the electrical hearing degradation. Therefore, it resulted in decreased spatial selective auditory attention. Increasing the target level to 75 dB SPL improved the acoustic hearing which resulted in listeners relying on acoustic hearing rather than electrical hearing.

To see the overall trend of spatial hearing ability according to the different target and masker azimuth, we also examined each individual CI listener result as shown in figures Figure 5 through Figure 11. As expected, the results showed higher speech recognition scores at the target azimuths that were closer to the side of amplification.

The group mean average for the electrophysiological results were consistent with the behavioral results. For NH listeners, although the EEG-based speech detection accuracy decreased by increasing the sound level, the effect of the sound level was not statistically significant. The effect of the sound level was significant for CIs in EH category, and consistent with their behavioral results. There was no significant effect of level on EEG-based speech detection accuracy for CIs in AH and EAH categories. We should mention that the number of trials included in the electrophysiological experiment was much less than the trials included in the behavioral test. Due to the time demanding nature of the EEG test, we had less repetition in trials for each condition. In addition, only the data from 3 subjects was included in the AH and EAH groups. Thus, we may argue that the data collected for these groups was not enough to show any significant effect of the sound level.

For both NH and CI listeners, the EEG-based speech detection accuracy at some speech levels and azimuths were around the chance level (33%) in electrophysiological experiment. One reason may be the difficulty of the task as subjects would require attend the target passage and suppress two competing maskers. It also may suggest that the auditory attention detection approach used in this study is not robust enough to detect the attended speech in presence of two competing speeches. In previous studies conducted to detect the attended speech through neural response [14], [23], [24], [35]–[38], the target sound was detected in two speakers cocktail party scenario (one speaker as a target and one speaker as a masker). In real world scenarios, NH listeners are able to focus on the target speech in the presence of several sources of noise. In the current paper, we examined the neural entrainment to the speech envelope in a three speakers cocktail party scenario. The linear model we used in this study was not able to detect the auditory attention at some of the sound levels due probably to considerable amount of maskers. We will continue to explore other methods, including non-linear models, to detect the selective auditory attention in presence of multiple maskers. To be able to generalize the results, we will need to collect more data with more subjects in each group (electrical, acoustic, electric+acoustic hearing).

## 5- Conclusion

This study examined the contribution of speech level on the spatial auditory selective attention in acoustic and electrical hearing. Both behavioral and physiological results showed that, at a fixed TMR, changing the level of the target speech in conversational levels range will not affect the selective auditory attention in NH listeners. However, our behavioral results for CI listeners with electrical hearing only showed a significant effect of speech level. Increasing the speech level resulted in decreased speech recognition score. In higher speech levels while keeping the target to masker ratio the same, both target and maskers' level increased. It implies that in electrical hearing, listeners prefer lower masking level at the cost of lower audibility of target. In hearing impaired listeners with acoustic hearing only, increasing the speech level resulted in increased speech recognition score. It means that in this case, having an audible target is more important than suppressing the maskers.

While the overall trend of electrophysiological results was consistent with the behavioral results, the speech detection accuracy in electrophysiological experiment was not high enough to justify its efficacy at some speech levels. Comparing to the other studies that examined the neural tracking of the target speech in presence of one competing sound, current study is more analogous to real life in which listeners have challenge to confront multiple sources of masking at the same time. Improving the neural detection accuracy may be possible with a nonlinear model. We will further examine the neural tracking of the speech envelope using a nonlinear model such as deep neural networks, which can learn the nonlinearities associated with the spatial selective auditory attention.

## 6- References

[1]     R. V Shannon, F. Zeng, V. Kamath, J. Wygonski, and M. Ekelid, "Speech Recognition with

Primarily Temporal Cues," *Science (80-. ).*, vol. 270, no. 5234, pp. 303–304, 1995.

[2]  B. G. Shinn-cunningham and V. Best, "Selective Attention in Normal and Impaired Hearing," *Trends Amplif.*, vol. 12, no. 4, pp. 283–299, 2008.

[3]  C. L. Mackersie, T. L. Prida, and D. Stiles, "The role of sequential stream segregation and frequency selectivity in the perception of simultaneous sentences by listeners with sensorineural hearing loss," *J. Speech, Lang. Hear. Res.*, vol. 44, pp. 19–28, 2001.

[4]  N. Marrone, C. R. Mason, and G. Kidd, "The effects of hearing loss and age on the benefit of spatial separation between multiple talkers in reverberant rooms," vol. 124, no. 5, pp. 3064–3075, 2008, doi: 10.1121/1.2980441.

[5]  V. Best, N. Marrone, C. R. Mason, G. Kidd, and B. G. Shinn-Cunningham, "Effects of Sensorineural Hearing Loss on Visually Guided Attention in a Multitalker Environment," *J. Assoc. Res. Otolaryngol.*, vol. 10, no. 1, pp. 142–149, 2008, doi: 10.1007/s10162-008-0146-7.

[6]  L. Dai, V. Best, and B. G. Shinn-cunningham, "Sensorineural hearing loss degrades behavioral and physiological measures of human spatial selective auditory attention," *Proc. Natl. Acad. Sci.*, pp. 1–10, 2018, doi: 10.1073/pnas.1721226115.

[7]  J. Murphy, A. Q. Summerfield, G. M. O. Donoghue, and D. R. Moore, "Spatial hearing of normally hearing and cochlear implanted children," *Int. J. Pediatr. Otorhinolaryngol.*, vol. 75, no. 4, pp. 489–494, 2011, doi: 10.1016/j.ijporl.2011.01.002.

[8]  M. Mok, L. Galvin, R. C. Dowell, and M. Mckay, "Spatial Unmasking and Binaural Advantage for Children with Normal Hearing , a Cochlear Implant and a Hearing Aid , and Bilateral Implants," *Audiol. Neurotol.*, vol. 12, no. 5, pp. 295–306, 2007, doi: 10.1159/000103210.

[9]  E. E. Bennett and R. Y. Litovsky, "Sound Localization in Toddlers with Normal Hearing and with Bilateral Cochlear Implants Revealed Through a Novel '"Reaching for Sound"' Task," *J. Am.*

*Acad. Audiol.*, 2019, doi: 10.3766/jaaa.18092.

[10] H. M. Wright, W. Bulla, and E. W. Tarr, "Spatial release from masking and sound localization using real-time sensorineural hearing loss and cochlear implant simulation," *J. Acoust. Soc. Am.*, vol. 145, no. 3, pp. 1877–1877, 2019.

[11] M. B. Winn, J. H. Won, and I. J. Moon, "Assessment of spectral and temporal resolution in cochlear implant users using psychoacoustic discrimination and speech cue categorization," *Ear Hear.*, vol. 37, no. 6, p. e377, 2016, doi: 10.1097/AUD.0000000000000328.Assessment.

[12] J. Seebacher, A. Franke-trieger, V. Weichbold, P. Zorowka, and K. Stephan, "Improved interaural timing of acoustic nerve stimulation affects sound localization in single-sided deaf cochlear implant users," *Hear. Res.*, vol. 371, pp. 19–27, 2019, doi: 10.1016/j.heares.2018.10.015.

[13] T. Francart, J. Brokx, and J. Wouters, "Sensitivity to Interaural Level Difference and Loudness Growth with Bilateral Bimodal Stimulation," *Audiol. Neurotol.*, vol. 13, no. 5, pp. 309–319, 2008, doi: 10.1159/000124279.

[14] T. L. Arbogast, C. R. Mason, G. Kidd, T. L. Arbogast, C. R. Mason, and G. Kidd, "The effect of spatial separation on informational and energetic masking of speech," *J. Acoust. Soc. Am.*, vol. 112, no. 5, pp. 2086–2098, 2002, doi: 10.1121/1.1510141.

[15] H. Glyde, L. Hickson, S. Cameron, and H. Dillon, "Problems Hearing in Noise in Older Adults: A Review of Spatial Processing Disorder," *Trends Amplif.*, vol. 15, no. 3, pp. 116–126, 2011, doi: 10.1177/1084713811424885.

[16] V. Best, N. Marrone, C. R. Mason, and G. Kidd, "The influence of non-spatial factors on measures of spatial release from masking," *J. Acoust. Soc. Am.*, vol. 131, no. 4, pp. 3103–3110, 2012, doi: 10.1121/1.3693656.

[17] O. Strelcyk and T. Dau, "Relations between frequency selectivity , temporal fine-structure

processing , and speech reception in impaired hearing," *J. Acoust. Soc. Am.*, vol. 125, no. 5, pp. 3328–3345, 2009, doi: 10.1121/1.3097469.

[18]  W. J. C. T. Smoski, "Discrimination of interaural temporal disparities by normal-hearing listeners and listeners with high-frequency sensorineural hearing loss," *J. Acoust. Soc. Am.*, vol. 79, no. 5, pp. 1541–1547, 1986.

[19]  D. B. Hawkins and F. L. Wightman, "Interaural Time Discrimination Ability of Listeners with Sensorineural Hearing Loss," *Audiology*, vol. 19, no. 6, pp. 495–507, 1980.

[20]  T. Y. C. Ching *et al.*, "Spatial release from masking in normal-hearing children and children who use hearing aids," *J. Acoust. Soc. Am.*, vol. 129, no. 1, pp. 368–375, 2011, doi: 10.1121/1.3523295.

[21]  F. J. Gallun, A. C. Diedesch, S. D. Kampel, K. M. Jakien, E. S. Sussman, and A. Einstein, "Independent impacts of age and hearing loss on spatial release in a complex auditory environment," *Front. Neurosci.*, vol. 7, no. December, pp. 1–11, 2013, doi: 10.3389/fnins.2013.00252.

[22]  S. Akbarzadeh, S. Lee, F. Chen, and C. Tan, "The effect of perceived sound quality of speech in noisy speech perception by normal hearing and hearing impaired listeners," *2019 41st Annu. Int. Conf. IEEE Eng. Med. Biol. Soc.*, pp. 3119–3122, 2019.

[23]  N. Mesgarani and E. F. Chang, "Selective cortical representation of attended speaker in multi-talker speech perception," *Nature*, vol. 485, no. 7397, p. 233, 2012, doi: 10.1038/nature11020.

[24]  J. A. O. Sullivan *et al.*, "Attentional Selection in a Cocktail Party Environment Can Be Decoded from Single-Trial EEG," *Cereb. Cortex*, vol. 25, no. 7, pp. 1697–1706, 2014, doi: 10.1093/cercor/bht355.

[25]  C. Horton, R. Srinivasan, and M. D'Zmura, "Envelope responses in single-trial EEG indicate

attended speaker in a 'cocktail party,'" *J. Neural Eng.*, vol. 11, no. 4, p. 046015, 2014, doi: 10.1088/1741-2560/11/4/046015.

[26] J. Vanthornhout, L. Decruy, J. Wouters, J. Z. Simon, and T. Francart, "Speech Intelligibility Predicted from Neural Entrainment of the Speech Envelope," *J. Assoc. Res. Otolaryngol.*, pp. 1–11, 2018, doi: 10.1007/s10162-018-0654-z.

[27] G. Kidd, V. Best, and C. R. Mason, "Listening to every other word : Examining the strength of linkage variables in forming streams of speech," *J. Acoust. Soc. Am.*, vol. 124, no. 6, pp. 3793–3802, 2008, doi: 10.1121/1.2998980.

[28] R. M. Cox and C. G. Genevieve C. Alexander, "Development of the Connected Speech Test (CST).pdf," *Ear Hear.*, vol. 8, no. 5, pp. 119S-126S, 1987.

[29] R. M. Cox and D. M. Mcdaniel, "Development of the Speech Intelligibility Rating (SIR) test for hearing aid comparisons," *J. Speech, Lang. Hear. Res.*, vol. 32, no. 2, pp. 347–352, 1989.

[30] M. J. Crosse, G. M. Di Liberto, A. Bednar, and E. C. Lalor, "The Multivariate Temporal Response Function ( mTRF ) Toolbox : A MATLAB Toolbox for Relating Neural Signals to Continuous Stimuli," *Front. Hum. Neurosci.*, vol. 10, p. 604, 2016, doi: 10.3389/fnhum.2016.00604.

[31] R. Oostenveld, P. Fries, E. Maris, and J. Schoffelen, "FieldTrip : Open Source Software for Advanced Analysis of MEG , EEG , and Invasive Electrophysiological Data," *Comput. Intell. Neurosci.*, vol. 2011, p. 1, 2011, doi: 10.1155/2011/156869.

[32] F. Zeng *et al.*, "Speech dynamic range and its effect on cochlear implant performance," *J. Acoust. Soc. Am.*, vol. 111, no. 1, pp. 377–386, 2002.

[33] J. B. Firszt *et al.*, "Recognition of Speech Presented at Soft to Loud Levels by Adult Cochlear Implant Recipients of Three Cochlear Implant Systems," *Ear Hear.*, vol. 25, no. 4, pp. 375–387, 2004, doi: 10.1097/01.AUD.0000134552.22205.EE.

[34] S. Akbarzadeh, S. Lee, S. Singh, and C. Tuan-tan, "Implication of speech level control in noise to sound quality judgement," in *2018 Asia-Pacific Signal and Information Processing Association Annual Summit and Conference (APSIPA ASC)*, 2018, pp. 388–392.

[35] C. Horton, R. Srinivasan, and M. D'Zmura, "Envelope responses in single-trial EEG indicate attended speaker in a 'cocktail party,'" *J. Neural Eng.*, 2014, doi: 10.1088/1741-2560/11/4/046015.

[36] W. Biesmans, N. Das, T. Francart, and A. Bertrand, "Auditory-Inspired Speech Envelope Extraction Methods for Improved EEG-Based Auditory Attention Detection in a Cocktail Party Scenario," *IEEE Trans. Neural Syst. Rehabil. Eng.*, vol. 25, no. 5, pp. 402–412, 2016.

[37] Y. Wang, J. Zhang, N. Ding, J. Zou, and H. Luo, "Prior Knowledge Guides Speech Segregation in Human Auditory Cortex," *Cereb. Cortex*, pp. 1–11, 2018, doi: 10.1093/cercor/bhy052.

[38] B. Mirkovic, S. Debener, M. Jaeger, and M. De Vos, "Decoding the attended speech stream with multi-channel EEG : implications for online , daily-life applications," *J. Neural Eng.*, vol. 12, no. 4, p. 046007, 2015, doi: 10.1088/1741-2560/12/4/046007.